# Study of parameters affecting the cooling capacity of liquid jets by using OpenFoam as tool to solve the inverse heat transfer problem

Kaissar Nabbout[1, *] and Martin Sommerfeld[1]

[1]Institute of Process Engineering, Otto-von-Guericke-University Magdeburg, 2 Universitätsplatz, 39106 Magdeburg, Germany

[*]Corresponding author: kaissar.deoliveira@ovgu.de

**Abstract**

In this work, some of the parameters influencing the cooling capacity of a liquid jet impinging onto Inconel 718 and C45 plates were experimentally investigated. The experiment included a high-speed camera to record the dynamic of the jet during the cooling process while an infrared camera was used to record the temperature field at the opposite surface. Jets made of water and oil-in-water emulsion were analysed as well as the influence of the oil concentration. Other parameters studied here include initial temperature of the plate, nozzle-to-plate distance, nozzle diameter, jet velocity, and impinging angle. The cooling performance was analysed by solving a full 3D inverse heat transfer problem (IHTP) with the Conjugate Gradient Method (CGM) implemented in a new solver in OpenFoam. The basic organization and implementation of the solver is shown, followed by its validation with a made-up case. Finally, the growth of the wetting front was analysed for different oil concentrations and a combination of nozzle diameters and jet velocities for the same flow rate. For the latest, unexpected results emerged when comparing the wetting growth for the two plate materials.

## 1. Introduction

Liquid jets are often used in applications requiring high cooling rates and precise temperature control, such as in nuclear power plant cooling, quenching, and machining processes. In the specific case of machining, cutting fluids are widely used to increase the tool lifetime and improve the surface quality of the workpiece. These results are achieved thanks to the fluid's ability to lubricate and cool the chip formation area, which is the region with highest heat flux due to friction and plastic deformation. The use of liquid jets with abundant quantity of cooling fluid during machining processes defines the so-called flood cooling strategy. This is probably the most popular cooling strategy used due to its effectiveness across a wide range of materials and processes. However, the enormous amount of coolant used has immense economical and environmental impact. Therefore, better understanding the parameters that have influence on the heat transfer between liquid jets and hot metals is essential to choose the best cooling strategy and make the process more efficient. For this reason, the influence of some parameters on the cooling performance of liquid jets are studied in this work. A new solver, implemented in OpenFoam, is proposed to solve the fully 3D Inverse Heat Transfer Problem (IHTP) of finding the space and time dependent heat flux boundary condition (BC) with multiple measurement positions for a plate cooled by a jet.

To make this work as clear as possible, it will be divided as follows: in the remaining of this first section, a brief introduction to jet cooling physics and commonly used methods to solve an IHTP are presented in two different subsections. The experimental setup and parameters studied are shown next on Section 2. The mathematical equations that constitute the solution of the IHTP are shown and explained on Section 3. Implementation and validation of the solver created in OpenFoam are briefly discussed in Section 4. The experimental and numerical results are presented and discussed on Section 5, followed by a conclusion and outlook of the work on Section 6.

*1.1. Jet cooling*

The cooling process by liquid jets is a very complex phenomenon that combines areas of knowledge that carry, by themselves, a high degree of complexity, such as the study of free jets and heat transfer mechanisms. When looking only at free jets, there are already many parameters that may influence their dynamics, as summarized by *Eggers and Villermaux* [1]. A change in the dynamic of the jet, affects, consequently, its cooling capacity and characteristics. One good exercise to see the influence of jet dynamics on cooling processes is to imagine two jets under identical conditions, each cooling identical plates under identical conditions. The only difference between them is that one jet is far enough from the plate, so it breaks up before touching it. It is intuitive to expect that the cooling process for one plate will be different from the other (which one is better or worse is not in discussion here). In processes involving high temperatures, the heat transfer mechanisms become much more complex due to phase change and boiling. There are three different boiling regimes, namely nucleate, transition and film boiling. All these regimes are usually present in jet cooling processes and can be also identified with help of a boiling curve, also know as Nukiyama curve, due to the pioneer work of *Nukiyama* [2]. In Figure 1, the main regions in jet cooling are shown and associated to their respective region in the Nukiyama curve.

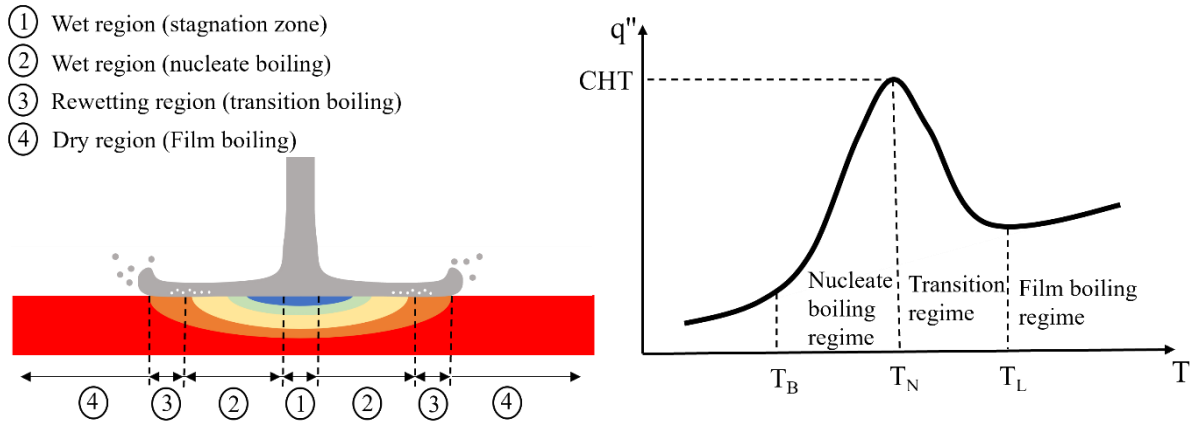

Figure 1: Illustration of the important regions for a liquid jet cooling a hot plate (left) and the association to these regions with the Nukiyama curve (right).

During the first moments when the liquid jet impinges onto the hot surface, the heat transfer between them can be very low if the plate has a temperature high enough to create a vapour layer that prevents direct contact between liquid and solid. Some fraction of second later, the liquid touches the surface and spreads forming a wet region, which grows with time due to jet's momentum. The boundary of the wet region is called rewetting front, and its growth is limited by the high temperatures of the dry region, which is still hot enough to sustain the vapour layer that prevents contact between liquid and solid. The temperature at which this vapour layer is formed is called Leidenfrost temperature, $T_L$, and coincides with the lowest heat flux region in the Nukiyama curve. The highest heat flux, or critical heat flux, *CHF*, is located somewhere in the rewetting region, which is the region in the neighbourhood of the rewetting front. The temperature related to the maximum heat flux is called Nukiyama temperature, $T_N$. Therefore, the *CHF* is found in a dynamic region that starts at the stagnation zone (region where the jet impacts and has almost zero velocity) and grows with the rewetting region.

The cooling effectiveness of liquid jets is usually evaluated based on heat flux, $q''$, or on the heat transfer coefficient, *h*. Another parameter often used is the Biot number, *Bi*, which is a dimensionless quantity that correlates the ratio of the conductive thermal resistance inside a body with the convective thermal resistance at the surfaces. Estimating *h* during real machining processes is a very challenging task and usually implies in limited instrumentation, being it contactless (pyrometers and infrared cameras) or not (thermocouples). In the work of *Kops and Arenson* [3], the authors found $h = 2500$ W/(m²K) for turning conditions on a lathe when using liquid coolant. The cooling capacity of six different cooling medias were evaluated by *Sales et al.* [4]. The authors also used dynamic tests under turning conditions and obtained an increased heat transfer coefficient for the emulsion with 10% concentration when compared to the one with 5%. Nevertheless, experiments to analyse the heat transfer capacity in jet cooling processes are usually not carried out with the machines used in the real process. Instead, facilities that mimic the conditions of the real process are used, which provides greater possibilities of instrumentation and a more controllable environment. Another advantage is the possibility to isolate and study parameters of interest. For instance, *Li et al.* [5] studied the heat flux, $q''$, of a plate being cooled by a water spray and the influence of the initial temperature of the plate. They analysed initial temperatures in the range from 400 °C up to 1000 °C and obtained higher heat fluxes for higher initial temperatures. Heat fluxes of the same order of magnitude were found by *Ciafalo et al.* [6] in their experiments involving the cooling of a hot plate by two water sprays impinging on the opposite sides of the plate. *Gradeck et al.* [7] analysed the influence on cooling when water is slightly polluted by oil. The authors obtained similar values of CHF for both jets (water and emulsion), however noticeable differences were observed for the heat transfer in the film boiling regime and for the velocity of the rewetting front. The influence of the jet Reynolds number, *Re*, on the growth of the rewetting front, and on the heat flux was studied by *Oliveira et al.* [8]. The authors observed that higher *Re* led to higher wetting front velocity and more substantial increase of $q''$ away from the impact location when compared to the stagnation zone. In jet cooling, not only the jet dynamics and fluid thermophysical properties affect the heat transfer process, but also the solid thermophysical properties and characteristics. *Mehdi et al.* [9] studied the heat transfer for a moving metal sheet cooled by two water sprays and noticed that the thicker sheets showed higher values for the heat flux. Also, three different materials were analysed: nickel, nicrofer and aluminium alloy AA6082. A conclusion about the role of the material properties on the results cannot be inferred, but it was clear that it has a direct influence on the heat flux. For instance, the nickel plate showed higher heat flux than the nicrofer plate for the same conditions. Many other parameters influence the heat transfer of jet cooling processes but will not be further discussed here. Despite the numerous works in this field, many aspects remain open, and hopefully they will, one day, be understood with the advent of new methods and techniques.

*1.2. Inverse heat transfer problems*

In most of the works previously mentioned, inverse heat transfer methods were used when either the heat flux $q''$ or the heat transfer coefficient *h* were needed. The reason why these methods have become so popular is because, first, they provide information about regions where direct access is not possible or extremely limited, and second, with modern computers they can provide the solution very quickly, sometimes even real-time solutions are possible. To better understand the concept of an IHTP, it is important to also define the classical direct heat transfer problem. In the direct problem, a body, with known thermal properties, is subject to known boundary conditions, which causes the body to have an unknown temperature distribution. Therefore, the direct problem is solved to determine the temperature distribution of the body. A change in boundary conditions, or in the material properties, results in a unique new temperature distribution within the body, making the direct problem a well-posed problem. Usually, in the IHTP, the temperature values are measured at specific locations of the body, and based on this measurements, boundary conditions or material properties can be estimated.

Two characteristics of IHTPs make them very challenging. First, the fact that there are multiple solutions that satisfy the same temperature measurements and second, the sensibility to measurement errors. For those reasons, inverse problems are classified as ill-posed and, therefore, need methods with a regularization (stabilization) technique to be solved. *Beck and Woodbury* [10] compared some of the most common methods used to solve IHTP: function specification (FS), Tikhonov regularization, conjugate gradient (CG) and singular value decomposition (SVD). Tikhonov method relies on the use of a regularization parameter that basically controls the trade-off between accuracy and stability. An analyse on the stability of Tikhonov regularization method can be found in the work of *Duda* [11]. The FS method, also called Beck's method, uses information about future time steps to estimate the information of interest, for example, a boundary condition. The number of time steps used is responsible to regularize the method and, therefore, must be appropriately chosen in order to balance the accuracy and stability of the solution. Two methods to compute the recommended number of forward time steps are proposed by *Komínek and Pohanka* [12]. Higher numbers of time steps make the solution more stable, but can drastically under predict high gradient peeks, while lower number of time steps can capture the high gradients but create very unstable solutions. Because of its characteristics of using only part of the measurements, the FS methods is classified as a sequential algorithm and has the advantage of being computationally cheap and to allow real-time analysis. When the complete duration of the experiment is used, the method is classified as whole-time domain algorithm. The CG method is a classic whole-time domain algorithm, although it is possible to create hybrid methods by combining it with sequential methods as shown in the work of *Lu et al.* [13]. Whole-time domain algorithms are generally more stable than sequential algorithms, but at the cost of more computational resources.

There are different types of the CG method as discussed by *Hào and Reinhardt* [14]. In this work, the CG method with adjoint problem for function estimation is used as proposed by *Özisik and Orlande* [15] and *Huang and Wang* [16]. Within the method, three problems are solved: the direct problem, the sensitivity problem, and the adjoint problem. The solution of these problems provides the parameters needed to iteratively estimate the heat flux, $q''$, during the jet cooling experiments. The influence of the procedure used to calculate the search direction used in the CG method is shown by *Colaço and Orlande* [17]. The authors compared three methods, namely Powell-Beale, Polak-Ribiere, and Fletcher-Revers, which is the one used in this work. The regularization of the inverse problem is done by choosing an adequate stopping criterion based on Morozov's discrepancy principle introduced by *Morozov* [18]. The principle states that the solution of the inverse problem is sufficiently accurate if the residual is proportional to the errors present in the measurements. Therefore, the regularization of the inverse problem lies in the number of iterations of the CG method. If the method stops too early or too late, the solution may not be accurate. The details and equations of the method are shown on Section 3.

## 2. Experimental setup and procedure

An illustration of the testbench facility used in the experiments is depicted in Figure 2a. A circular plate, with diameter $d_{plate}$ = 140 mm, made either of Inconel 718 or steel C45 is heated by and induction heater up to temperatures higher than the initial temperature, $T_{ini}$, of interest.

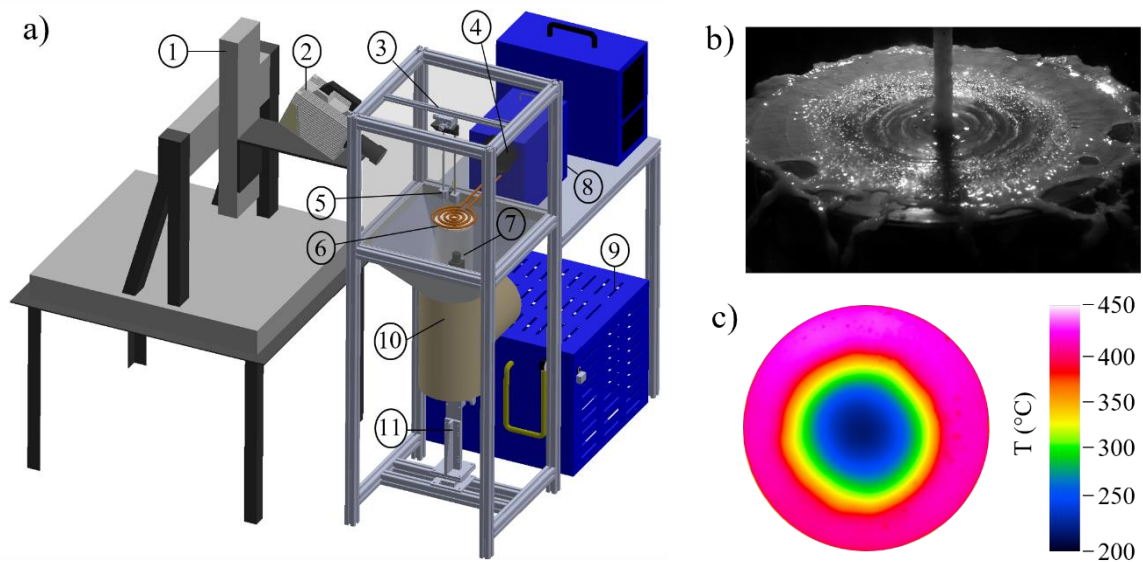

Figure 2: Illustration of the facility (a) and its components: 1) Traverse system for the high-speed camera; 2) High-speed camera; 3) Traverse system for the nozzle; 4) LED with heatsink; 5) Nozzle; 6) Plate; 7) Infrared camera; 8) Induction heating system; 9) Induction heater's cooling system; 10) Part of pipeline for the exhaustion system; 11) Traverse system for the infrared camera. Moment captured by the high-speed (b) and infrared (c) cameras during the cooling process.

Then, the induction heater is moved aside, and the plate is let to cool down naturally in order to have an initial temperature distribution as uniform as possible with values close to $T_{ini}$. Moments before the plate reaches the desired initial state, the nozzle is positioned at the desired height, $h_{nozzle}$, by a linear rail system controlled by an Arduino board. The impingement and evolution of the jet on the top surface is recorded by a Photron FASTCAM SA4 high speed camera with a framerate $f_{HS}$ = 1000 fps. The illumination is done with a Cree CXB3590 LED. The camera is positioned laterally to the jet with a measured inclination. An Infratec PIR uc 605 infrared camera is positioned under the plate to record the temperature field of the bottom surface with a framerate of $f_{IR}$ = 25 fps. Example of the results captured by the high-speed and infrared cameras are shown in Figure 2b and Figure 2c, respectively. The thermophysical properties of the plates and liquids used in the experiments are listed on Table 1. The values are for room temperature, approximately $T_{env}$ = 25 °C, and are considered constant.

Table 1: Thermophysical properties for the plates and liquids used.

| Solid | Density kg/m³ | Thermal conductivity W/m.K | Specific heat capacity J/Kg.K | Thermal diffusivity mm²/s |
|---|---|---|---|---|
| Inconel 718 | 8190 | 11.4 | 435 | 3.20 |
| C45 | 7800 | 45 | 480 | 12.02 |
| **Liquid** | **Density kg/m³** | **Surface tension mN/m** | **Specific heat capacity J/Kg.K** | **Dynamic viscosity mPa.s** |
| Water | 995.9 | 71.2 | 4181 | 0.805 |
| Adrana AY 401 | 962.2 | 28.1 | - | 68.084 |
| Water + 4% Adrana AY 401 | 993.9 | 33.7 | 3939 | 0.950 |
| Water + 8% Adrana AY 401 | 993.5 | 33 | 3906 | 1.022 |
| Water + 14% Adrana AY 401 | 993 | 32.7 | - | 1.233 |
| Water + 20% Adrana AY 401 | 992.3 | 32.7 | - | 1.576 |

The experiments conducted for this work, seeks to answer the following questions regarding the heat transfer in liquid jet cooling:

- What is the influence of oil concentration?
- What is the influence of the initial temperature, $T_{ini}$, of the plate?
- What is the influence of nozzle-to-plate distance, $H$?
- What is the influence of nozzle diameter, $d_{nozzle}$, when using the same flowrate?
- What is the influence of jet velocity, $V_{jet}$, when using the same $d_{nozzle}$?
- What is the influence $C$ and $V_{jet}$ on the growth of the wet region?
- What is the influence of impinging angle, $\theta_{jet}$?
- What is the influence of the plate material?
- Is Re a suitable parameter to be linked to cooling capacity of a jet?

In order to answer so many questions, numerous different experiments were conducted, and their parameters are listed on Table 2.

Table 2: Setups used in the experiments to analyse the influence of different parameters.

| Parameter of interest | d mm | H mm | T °C | C % | V m/s | $\theta_{jet}$ ° | Q l/min | Re |
|---|---|---|---|---|---|---|---|---|
| C | 7 | 10 | 500 | 0 | 2 | 90 | 4.7 | 17320 |
|   |   |    |     | 4 |   |    |     | 14647 |
|   |   |    |     | 8 |   |    |     | 13610 |
|   |   |    |     | 14|   |    |     | 11275 |
|   |   |    |     | 20|   |    |     | 8815  |
| Tini | 7 | 10 | 300 | 8 | 2 | 90 | 4.7 | 13610 |
|      |   |    | 400 |   |   |    |     |       |
|      |   |    | 500 |   |   |    |     |       |
| H | 7 | 10 | 500 | 8 | 2 | 90 | 4.7 | 13610 |
|   |   | 40 |     |   |   |    |     |       |
|   |   | 70 |     |   |   |    |     |       |
| $d_{nozzle}$ | 3 | 10 | 500 | 8 | 11 | 90 | 4.7 | 32080 |
|              | 5 |    |     |   | 4  |    |     | 19442 |
|              | 7 |    |     |   | 2  |    |     | 13610 |
| $V_{jet}$ | 7 | 10 | 500 | 8 | 2   | 90 | 4.7  | 13610 |
|           |   |    |     |   | 2.9 |    | 6.7  | 19734 |
|           |   |    |     |   | 4.7 |    | 10.9 | 31983 |
| $\theta_{jet}$ | 7 | 10 | 500 | 8 | 2 | 30 | 4.7 | 13610 |
|                |   |    |     |   |   | 60 |     |       |
|                |   |    |     |   |   | 90 |     |       |

## 3. Mathematical formulation of the inverse Problem and CG method

The problem to be solved here is illustrated in Figure 3. It consists in the estimation of the transient and space-dependent heat flux, $q''(S_{top},t)$, on the top surface, $S_{top}$, which is impinged by a liquid jet. Because direct measurements on $S_{top}$ are not possible, the temperature measurement, $T_{meas}(P_{meas},t)$, is taken at specific locations, $P_{meas}$, on $S_{bottom}$ in order to solve an IHTP. The solution of the IHTP comprise the minimization of the objective function:

$$J[q''(S_{top},t)] = \int_{t=0}^{t_f} [T_{meas}(P_{meas},t) - T_{sim}(P_{meas},t)]^2 dt, \tag{1}$$

in which $t_f$ is the analysed duration of the experiment and $T_{sim}(P_{meas},t)$ are the temperatures at $P_{meas}$ obtained by the solution of the direct problem when the plate is subjected to $q''(S_{top},t)$.

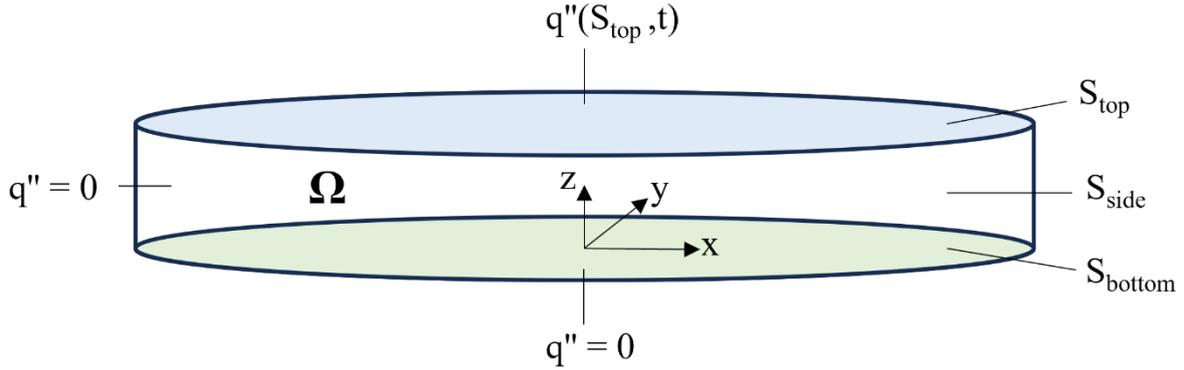

Figure 3: Illustration of the problem to be solved and its boundary conditions.

### 3.1. Direct problem

The solution of the direct problem provides the temperature field, $T(x,y,z,t)$, in the plate when $q''(S_{top},t)$ is known. The other surfaces are considered isolated, because the heat flux on the top surface is considered orders of magnitude greater. Therefore, the problem can be written as:

$$\frac{\partial T}{\partial t} - \alpha \left( \frac{\partial^2 T}{\partial x^2} + \frac{\partial^2 T}{\partial y^2} + \frac{\partial^2 T}{\partial z^2} \right) = 0 \quad \text{in } \Omega, \quad t > 0 \tag{2a}$$

$$\frac{\partial T}{\partial n} = 0 \qquad \text{on } S_{side} \text{ and } S_{bottom}, \quad t > 0 \qquad (2b)$$

$$\frac{\partial T}{\partial n} = q''(S_{top}, t) \qquad \text{on } S_{top}, \quad t > 0 \qquad (2c)$$

$$T = T_{ini} \qquad \text{in } \Omega, \quad t = 0 \qquad (2d)$$

### 3.2. Sensitivity problem

The sensitivity problem needs to be solved in order to compute the search step size, $\beta$, of the CG method. The basic idea behind the sensitivity problem is that the temperature field $T(x,y,z,t)$ changes by an amount $\Delta T(x,y,z,t)$, when the heat flux $q''(S_{top},t)$ undergoes a variation $\Delta q''(S_{top},t)$. Therefore, replacing $[T]$ by $[T + \Delta T]$ and $[q'']$ by $[q'' + \Delta q'']$ in Eq. (2 and subtracting the resulting equations from the original ones, the sensitivity problem is then written as:

$$\frac{\partial \Delta T}{\partial t} - \alpha \left( \frac{\partial^2 \Delta T}{\partial x^2} + \frac{\partial^2 \Delta T}{\partial y^2} + \frac{\partial^2 \Delta T}{\partial z^2} \right) = 0 \qquad \text{in } \Omega, \quad t > 0 \qquad (3a)$$

$$\frac{\partial \Delta T}{\partial n} = 0 \qquad \text{on } S_{side} \text{ and } S_{bottom}, \quad t > 0 \qquad (3b)$$

$$\frac{\partial \Delta T}{\partial n} = \Delta q''(S_{top}, t) \qquad \text{on } S_{top}, \quad t > 0 \qquad (3c)$$

$$T = 0 \qquad \text{in } \Omega, \quad t = 0 \qquad (3d)$$

### 3.3. Adjoint problem

The derivation of the adjoint problem is similar to the sensitivity problem, although much more laborious. The idea here is to minimize the objective function given by Eq. (*1* by using a Lagrange multiplier, $\lambda(x,y,z,t)$. The procedure starts by multiplying the direct problem, Eq. (*2*a, by $\lambda$ and then integrating the expression over time and space. The resulting expression is added to Eq. (*1*, which gives the following modified equation of the object functional:

$$J = \int_{t=0}^{t_f} [T_{meas} - T_{sim}]^2 dt + \int_{\Omega} \int_{t=0}^{t_f} \lambda \left[ \frac{\partial T}{\partial t} - \alpha \left( \frac{\partial^2 T}{\partial x^2} + \frac{\partial^2 T}{\partial y^2} + \frac{\partial^2 T}{\partial z^2} \right) \right]^2 dt\, d\Omega. \qquad (4)$$

The variation $\Delta J[q''(S_{top},t)]$ of the function $J[q''(S_{top},t)]$ is obtained by assuming that $T(x,y,z,t)$ is perturbed by $\Delta T(x,y,z,t)$ when $q''(S_{top},t)$ is perturbed by $\Delta q''(S_{top},t)$. So, similarly to the sensitivity problem, the following expression is obtained when replacing $[T]$ by $[T + \Delta T]$, [q''] by [q''+ Δq''] and [J] by [J+ ΔJ] on Eq. (4, then subtracting from the resulting expression the original equation, and neglecting the second order terms:

$$\Delta J = \int_{t=0}^{t_f} 2(T_{sim} - T_{meas})\Delta T_{meas} dt \\ + \int_{\Omega} \int_{t=0}^{t_f} \lambda \left[\frac{\partial \Delta T}{\partial t} - \alpha \left(\frac{\partial^2 \Delta T}{\partial x^2} + \frac{\partial^2 \Delta T}{\partial y^2} + \frac{\partial^2 \Delta T}{\partial z^2}\right)\right]^2 dt\, d\Omega. \quad (5)$$

The second term on the right-hand side can be separated in four integrals and be solved applying integration by parts and the initial conditions of the sensitivity problem, resulting in:

$$\int_{t_0}^{t_f} \lambda \frac{\partial \Delta T}{\partial t} dt = \lambda(\Omega, t_f).\Delta T(\Omega, t_f) - \int_{t_0}^{t_f} \Delta T \frac{\partial \lambda}{\partial t} dt \quad (6a)$$

$$\int_{x_0}^{x_f} \lambda \frac{\partial^2 \Delta T}{\partial x^2} dx = \left[\frac{\partial \lambda}{\partial x}\Delta T\right]_{x_0} - \left[\frac{\partial \lambda}{\partial x}\Delta T\right]_{x_f} + \int_{x_0}^{x_f} \Delta T \frac{\partial^2 \lambda}{\partial x^2} dx \quad (6b)$$

$$\int_{y_0}^{y_f} \lambda \frac{\partial^2 \Delta T}{\partial y^2} dy = \left[\frac{\partial \lambda}{\partial y}\Delta T\right]_{y_0} - \left[\frac{\partial \lambda}{\partial y}\Delta T\right]_{y_f} + \int_{y_0}^{y_f} \Delta T \frac{\partial^2 \lambda}{\partial y^2} dy \quad (6c)$$

$$\int_{z_0}^{z_f} \lambda \frac{\partial^2 \Delta T}{\partial z^2} dz = \left[\frac{\partial \lambda}{\partial z}\Delta T\right]_{z_0} - \left[\frac{\partial \lambda}{\partial z}\Delta T\right]_{z_f} - [\lambda.\Delta q'']_{z_f} + \int_{z_0}^{z_f} \Delta T \frac{\partial^2 \lambda}{\partial z^2} dy \quad (6d)$$

When replacing back Eq. (6 into Eq. (5, the adjoint problem is obtained by allowing all terms containing ΔT to go to zero. Therefore, the last terms on the right-hand side of Eq. (6 constitute the main equation of the adjoint problem while the other terms (containing ΔT) are either the boundary conditions or the final value of the problem. Because the measurements are taken on a boundary (z = 0), the first term on the right-hand side of Eq. (5 is also added as boundary condition. Therefore, the adjoint problem can be written as:

$$\frac{\partial \lambda}{\partial t} + \alpha \left(\frac{\partial^2 \lambda}{\partial x^2} + \frac{\partial^2 \lambda}{\partial y^2} + \frac{\partial^2 \lambda}{\partial z^2}\right) = 0 \quad \text{in } \Omega, \quad t > 0 \quad (7a)$$

$$\frac{\partial \lambda}{\partial n} = 0 \qquad \text{on } S_{side} \text{ and } S_{top}, \quad t > 0 \qquad (7b)$$

$$\frac{\partial \lambda}{\partial n} = \frac{2(T_{sim} - T_{meas})\delta_{meas}}{\alpha} \qquad \text{on } S_{bottom}, \quad t > 0 \qquad (7c)$$

$$\lambda = 0 \qquad \text{in } \Omega, \quad t = t_f \qquad (7d)$$

The adjoint problem obtained bellow is a final value problem and can easily be converted into an initial value problem by replacing the time variable to $\tau = t_f - t$. The problem is then rewritten as:

$$\frac{\partial \lambda}{\partial \tau} - \alpha \left( \frac{\partial^2 \lambda}{\partial x^2} + \frac{\partial^2 \lambda}{\partial y^2} + \frac{\partial^2 \lambda}{\partial z^2} \right) = 0 \qquad \text{in } \Omega, \quad \tau < t_f \qquad (8a)$$

$$\frac{\partial \lambda}{\partial n} = 0 \qquad \text{on } S_{side} \text{ and } S_{top}, \quad \tau < t_f \qquad (8b)$$

$$\frac{\partial \lambda}{\partial n} = -\frac{2(T_{sim} - T_{meas})\delta_{meas}}{\alpha} \qquad \text{on } S_{bottom}, \quad \tau < t_f \qquad (8c)$$

$$\lambda = 0 \qquad \text{in } \Omega, \quad \tau = 0 \qquad (8d)$$

Here $\delta_{meas}$ is the Dirac delta function. Finally, the only term left is the one that does not contain $\Delta T$:

$$\Delta J = \int_{S_{top}} \int_{t=0}^{t_f} \alpha \cdot \lambda(S_{top}, t) \cdot \Delta q''(S_{top}, t) \, dt \, dS_{top} \qquad (9)$$

The variation $\Delta J(S_{top}, t)$ gives the directional derivative of $J(S_{top}, t)$ in the direction of the perturbation $\Delta q''(S_{top}, t)$. Therefore, by definition, it can be written as:

$$\Delta J = \int_{S_{top}} \int_{t=0}^{t_f} \nabla J[q''(S_{top}, t)] \cdot \Delta q''(S_{top}, t) \, dt \, dS_{top} \qquad (10)$$

By comparing Eq. (*9* and Eq. (*10* the following expression is obtained for the gradient equation of the function *J*.

$$\nabla J[q''(S_{top},t)] = \alpha.\lambda(S_{top},t) \tag{11}$$

From Eq. (*8d* it becomes clear that the gradient $\nabla J$ is always equal to zero at the final time $t_f$, which can lead to wrong calculations of $q''$ close to the final time. To overcome this problem, the gradient at the final time step is considered equal to the gradient of the previous time step.

### 3.4. Conjugate gradient method

As previously mentioned, the estimation of the heat flux $q''(S_{top},t)$ is done by minimizing Eq. (*1*. This is achieved by iteratively estimating a new heat flux based on the CG method given by:

$$q''^{k+1}(S_{top},t) = q''^{k}(S_{top},t) + \beta^k d^k(S_{top},t) \tag{12}$$

where $\beta^k$ is the search step size and $d^k(S_{top},t)$ is the search direction, defined by the conjugation of the gradient at iteration *k* and the search direction at iteration *k-1*:

$$d^k(S_{top},t) = -\nabla J[q''^{k}(S_{top},t)] + \gamma^k d^{k-1}(S_{top},t) \tag{13}$$

In the equation above, $\gamma^k$ is the conjugation coefficient and is calculated based on Fletcher-Reeves method [19]:

$$\gamma^k = \frac{\int_{S_{top}} \int_{t=0}^{t_f} \{\nabla J[q''^{k}(S_{top},t)]\}^2 \, dt \, dS_{top}}{\int_{S_{top}} \int_{t=0}^{t_f} \{\nabla J[q''^{k-1}(S_{top},t)]\}^2 \, dt \, dS_{top}} \tag{14}$$

The search step size is calculated accord to the following equation:

$$\beta^k = \frac{\int_{t=0}^{t_f} \sum_1^M [T_{sim}(P_{meas},t) - T_{meas}(P_{meas},t)].\Delta T(P_{meas},t) \, dt}{\int_{t=0}^{t_f} \sum_1^M [\Delta T(P_{meas},t)]^2 \, dt} \tag{15}$$

where M is the number of measured positions in the experiments.

*3.5. Stopping criterion*

As already mentioned, the discrepancy principle is used as stopping criterion and is responsible for the regularization of the IHTP. The principle states that the solution is sufficiently accurate if the error in the temperature estimation from the solution of the IHTP at *P<sub>meas</sub>* is the same order of magnitude that the standard deviation of the measurement errors. This can be written as:

$$|T_{meas}(P_{meas},t) - T_{sim}(P_{meas},t)| \approx \sigma[T_{meas}(P_{meas},t)] \tag{16}$$

where $\sigma[T_{meas}(S_{bottom},t)]$ is the standard deviation of the temperature measurements at the bottom surface. In the case of the PIR uc 605 infrared camera, the measurement error provided by the manufacturer is 5%, meaning that the value for $\sigma$ at each measured position is given by the camera error multiplied by the measured temperature at that location. The stopping criterion is therefore calculated by:

$$J[q''(S_{top},t)] < \int_{t=0}^{t_f} \sum_{1}^{M} [\sigma(T_{meas})]^2 \, dt \tag{17}$$

With the calculation of the stopping criterion, the mathematical formulation for the solution of the IHTP with the CG method for function estimation is completed. A histogram with the algorithm for the iterative procedure is presented in the next section.

## 4. Implementation in OpenFoam and validation

As a mature free open-source software, OpenFoam gives the possibility to modify any part of the code, create new tools or solvers as needed, and share it with its ever-growing community. Those were the main reasons why OpenFoam was chosen as the tool to solve the IHTP in this work. For better understanding, a brief explanation of OpenFoam's folder organization is given with help of Figure 4. The *0* folder contains the initial fields and BC for all variables: *T* for the temperature field of the direct problem, *sensiT* for the *ΔT* field of the sensitivity problem and *LM* for the Lagrange multiplier *λ* of the adjoint problem. The *system* folder contains the files associated with mesh generation and parameters control (time step, discretisation schemes, tolerances, etc.). Information about the mesh, physical properties, and stopping criterion for the CG method can be found in the *constant* folder.

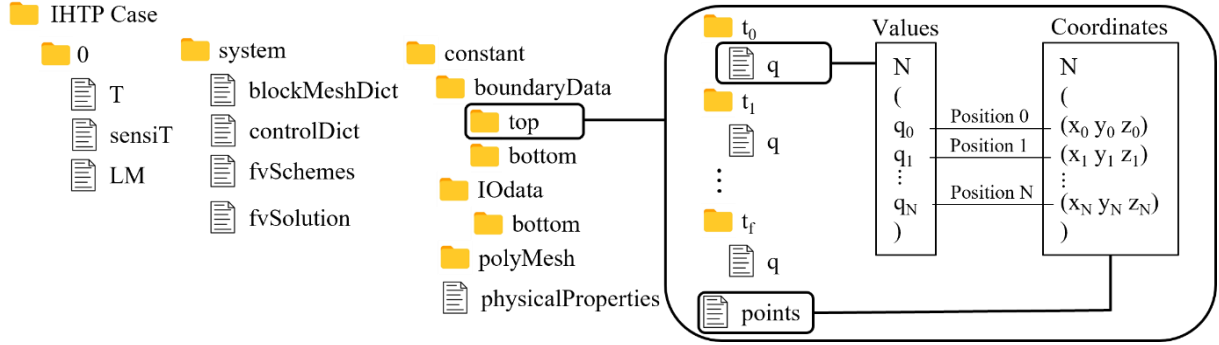

Figure 4: Folder structure used by the *ihtFoam* solver.

The development of the new solver can be separated into two steps: implementation of CG method and creation of a new boundary condition for the transient and space-dependent heat flux. The new BC is a modification of the already existent *timeVaryingMappedFixedValue*, which interpolates values from a set of supplied points in space and time, and then apply these values as Dirichlet boundary condition. The new boundary condition uses the same interpolation functionalities, but applies the values as Neumann boundary condition, therefore it is called *timeVaryingMappedFixedGradient*.

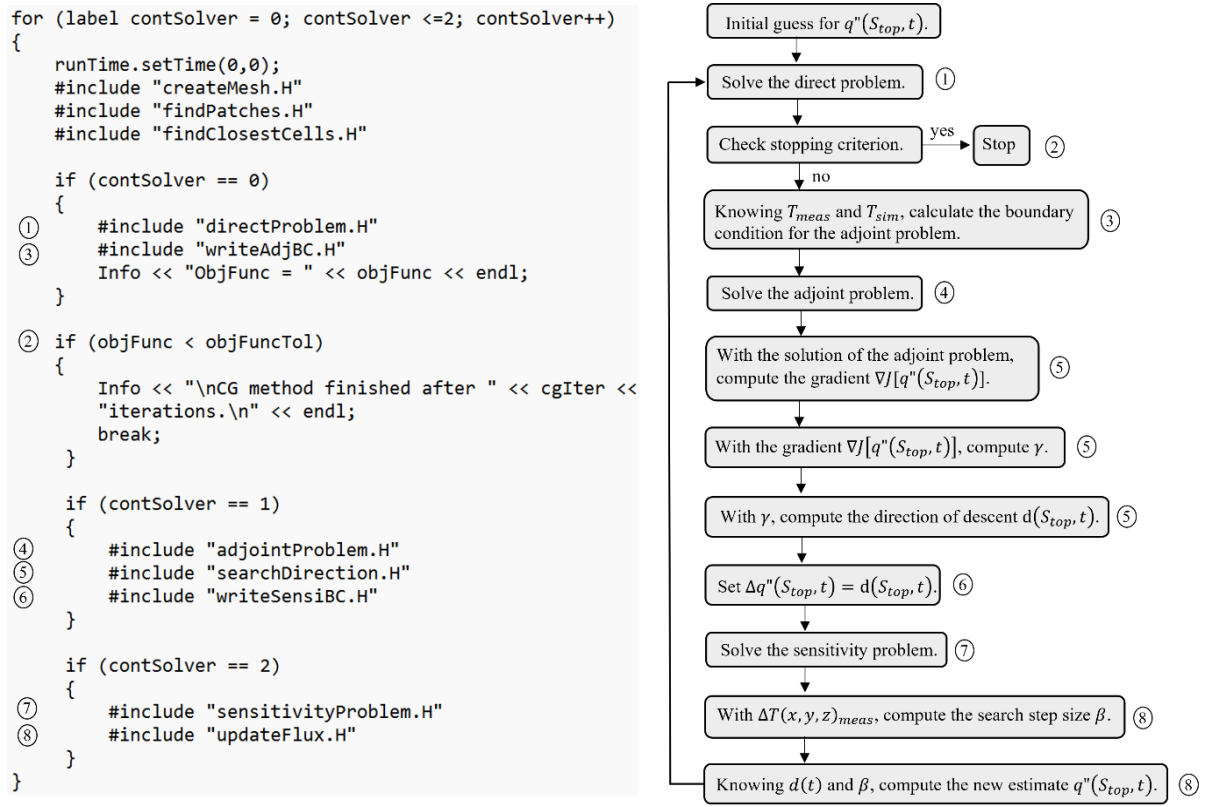

Figure 5: Association of the core part of the *ihtFoam* (left) and the steps of the CG method represented in a histogram (right).

It is important to highlight that the gradient value is applied as boundary condition and not directly the heat flux, which can be obtained multiplying the gradient by the heat conductivity. Detailed description of the code implementation is not provided here, but rather a brief clarification of how the steps of the CG method is implemented in the new solver *ihtFoam*. For that purpose, the main part of the code is shown in Figure 5 together with a histogram that summarizing the CG method. The code is an extension of the solver *laplacianFoam* to solve not only the direct problem, but also the adjoint and sensitivity problems. The code lines on the left are associated to the histogram by numbers from 1 to 8, highlighting what is done inside each header (.H) file.

*4.1. Solver validation*

To validate the code, a test case like the one proposed by *Huang and Wang* [16] is used. The idea is to check if the solver is able to recover the heat flux on the top surface of a square plate with size 12x12x1.2 m³ and grid 12x12x6 cells, in x, y, and z directions, respectively. The imposed heat flux is given by:

$$q_1''(I,J,t) = 60 \sin\left(\frac{t}{t_f}\pi\right), \quad \text{for } 1 \leq I \leq 12, 1 \leq J \leq 12, \text{ and } 0 \leq t \leq 24 \tag{18a}$$

$$q_2''(I,J,t) = 60 \sin\left(\frac{t}{t_f}\pi\right), \quad \text{for } 3 \leq I \leq 10, 3 \leq J \leq 10, \text{ and } 0 \leq t \leq 24 \tag{18b}$$

$$q_3''(I,J,t) = 40 \sin\left(\frac{t}{t_f}\pi\right), \quad \text{for } 5 \leq I \leq 8, 5 \leq J \leq 8, \text{ and } 0 \leq t \leq 24 \tag{18c}$$

$$q_{imposed}''(I,J,t) = q_1''(I,J,t) + q_2''(I,J,t) + q_3''(I,J,t) \quad \text{in } \Omega, \quad 0 \leq t \leq 24 \tag{18d}$$

where I and J are the grid index for x and y directions, respectively. All other surfaces are considered insulated. The heat flux is estimated at the centre of each cell at the top surface (144 cells) for each time step (24 seconds). The input to solve the inverse problem are the temperatures of the cells at the bottom surface, meaning 144 "measurement" points. These temperatures were obtained by solving the direct problem and extracting the values for each cell at each time step.

The distribution of $q_{imposed}''$ (left) and $q_{ihtFoam}''$ (right) are shown in Figure 6a for a time $t = 12s$. It is noticeable that the estimated heat flux cannot capture the drastic changes between the three regions and generates smeared results in the transition region. For a more quantitative analysis, the heat flux profile over all cells at J = 7 (y = 6.5 m) is shown in Figure 6b for times

t = 6s and t = 12s. The values at the central region of the plate were well captured and presented smaller errors when compared to the values outside the central region due to the jumps in heat flux when changing regions.

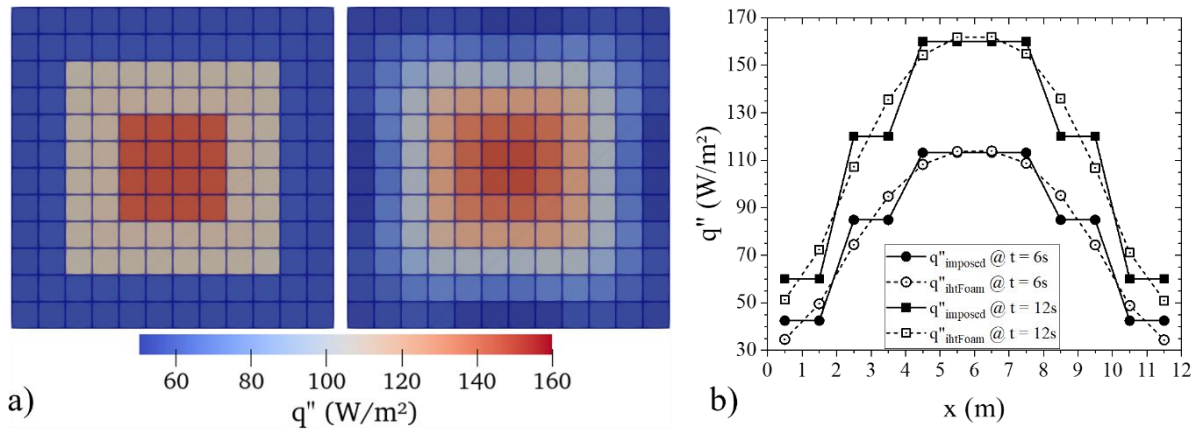

Figure 6: a) Distribution for the imposed heat flux (left) and the heat flux estimated by the *ihtFoam* (right) for a time t = 12s; b) Profile over the cells located at J = 7 (y = 6.5 m) at times t = 6s and t = 12s.

## 5. Results and discussion

The cooling capacity of the different experimental setups showed on Table 2 is compared based on the heat flux calculated by the *ihtFoam* solver. The values of heat flux shown in the following subsections are for the impinging region at the top surface, except for the last subsection in which the influence of the distance from the impinging zone is analysed. The parameters of interest are oil concentration ($C$), initial plate temperature ($T_{ini}$), nozzle-to-plate distance ($H$), nozzle diameter ($d$), jet velocity ($V_{jet}$), and impinging angle ($\theta_{jet}$). For each parameter, the influence of plate material is shown for Inconel 718 and steel C45. The plates have diameter $d_{plate}$ = 140 mm and thickness $h_{plate}$ = 5 mm. As input for the simulations, 249 equally spaced measurement points were used at the bottom surface, covering a circular region of diameter 90 mm, as shown in Figure 7a. For the estimation of the gradient, and therefore the heat flux, at the top surface, 307 equally spaced points (over the whole top surface) were used for the boundary condition *timeVaryingMappedFixedGradient*. The mesh used was the same for all cases, consisting of approximately 17,000 cells with approximate size of 2x2x1.25 mm, as shown in Figure 7b.

Before proceeding to the results, a brief discussion on the initial conditions is necessary. The initial temperature distribution of the plate is considered uniform and equal to the temperature at the centre of the bottom surface. However, the measured points include data for approximately 4 seconds before the jet is initiated. This interval safely covers the time needed by the bottom surface to detect the cooling at the top surface. It also provides a better initial temperature distribution of the plate before the cooling process initiates. The temperature distribution at the bottom surface, moments before the cooling process initiates, can be seen in Figure 8 for the experiment (left) and for the simulation (right). The use of this small time interval increases the accuracy of the solution, since it is clear that the initial temperature field in the real process is not uniform.

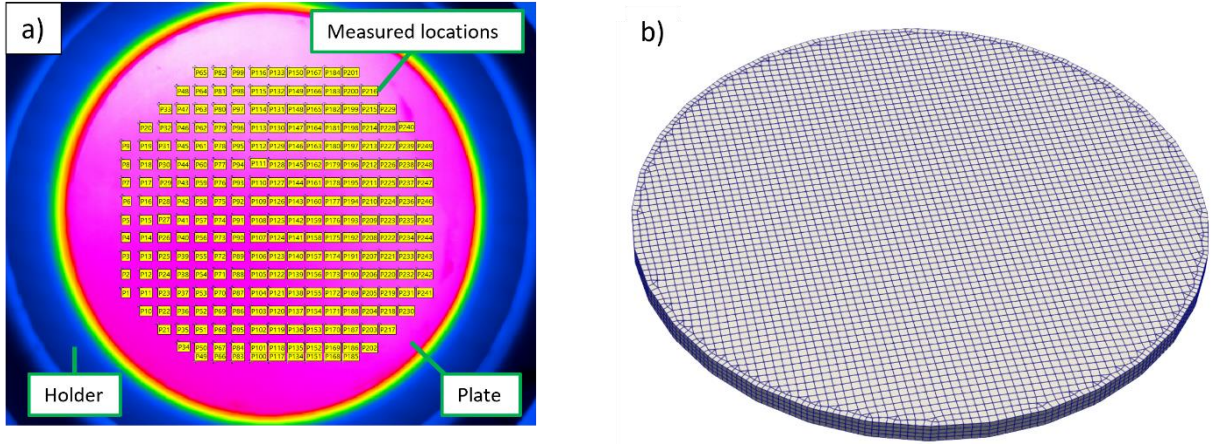

Figure 7: a) Image from the infrared camera showing the position of the 249 points used has input to the IHTP; b) Mesh used to solve the IHTP.

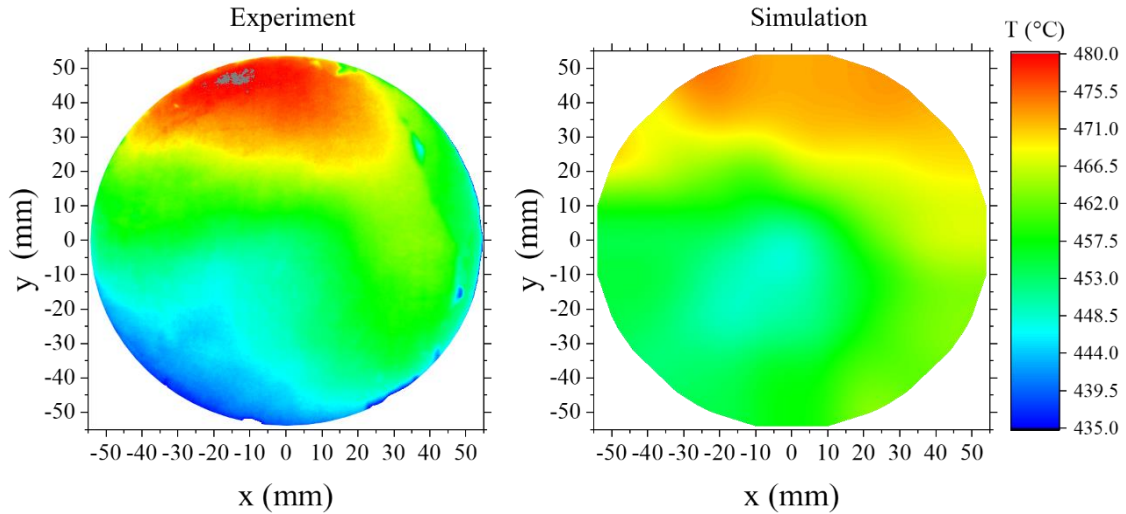

Figure 8: Temperature distribution at the bottom surface moments before the cooling process initiates.

## 5.1. *Influence of oil concentration*

The liquid studied here is an oil-in-water emulsion consisting of different concentrations of the mineral oil Adrana AY 401. This emulsion is commonly used in machining processes and the manufacturer recommends concentrations between 4% and 10% depending on the material to be machined. Therefore, initially 3 concentrations were analysed, $C = 0\%$ which means pure water, $C = 4\%$, and $C = 8\%$. As shown in the results from Figure 9, pure water presented the lowest values of heat flux for both Inconel 718 and C45. Interestingly, the highest heat flux was obtained by $C = 4\%$, with $C = 8\%$ being in between. This behaviour was more evident for Inconel 718. These results suggests that a small amount of oil can indeed enhance the cooling capacity of the emulsion. To verify the trend when increasing the concentration, $C = 14\%$ and $C = 20\%$ were also tested for C45, confirming that the increase in oil concentration tends to diminish the cooling capability of the liquid. It is important to highlight that during machining processes, the liquid is not only responsible for cooling, but also for lubricating the cutting zone. Therefore, the higher oil concentration could be advantageous for lubrication, which could also reduce temperatures in the process due to reduction of friction. It is also noticeable that the heat flux was higher for C45 when compared to Inconel 718. That is due to the higher thermal effusivity of C45, i.e., its ability to exchange heat with the environment. Interestingly, the ratio between the maximum heat flux for C45 and for Inconel 718 is approximately 2, which is also the ratio of the thermal effusivity of the materials. This observation is valid for all parameters studied and will no longer be mentioned.

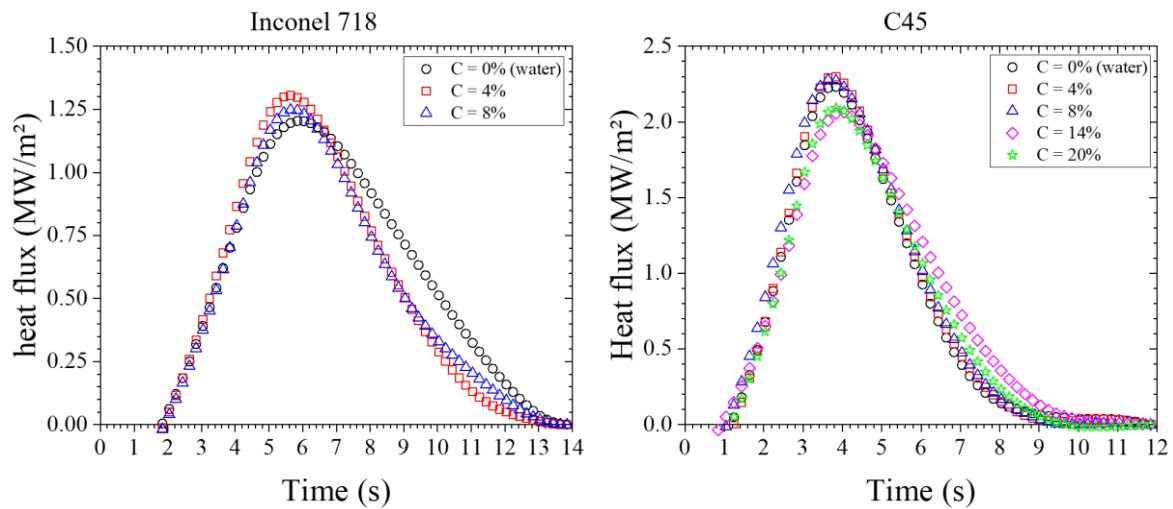

Figure 9: Influence of oil concentration on the heat flux for Inconel 718 (left) and C45 (right).

*5.2. Influence of initial temperature of the plate*

The initial temperature of the plate also affects the heat flux as shown in Figure 10 for $C = 8\%$ (as it will be used from now on). Three different values of $T_{ini}$ were analysed. As mentioned before, this value stands for the approximate temperature at the centre of the bottom surface and is not uniform over the plate. It is interesting to notice that the maximum heat flux for $T_{ini} = 450$ °C was approximately two times higher when compared to $T_{ini} = 250$ °C for both plates. This suggests that $T_{ini}$ may have the same influence on the heat flux independent of the material, however no conclusion can be inferred from analysing only two materials. The temperatures analysed here are not high enough to identify any boiling regime during the cooling process, as it was identified by *Li et al.* [5] for temperatures above 700°C.

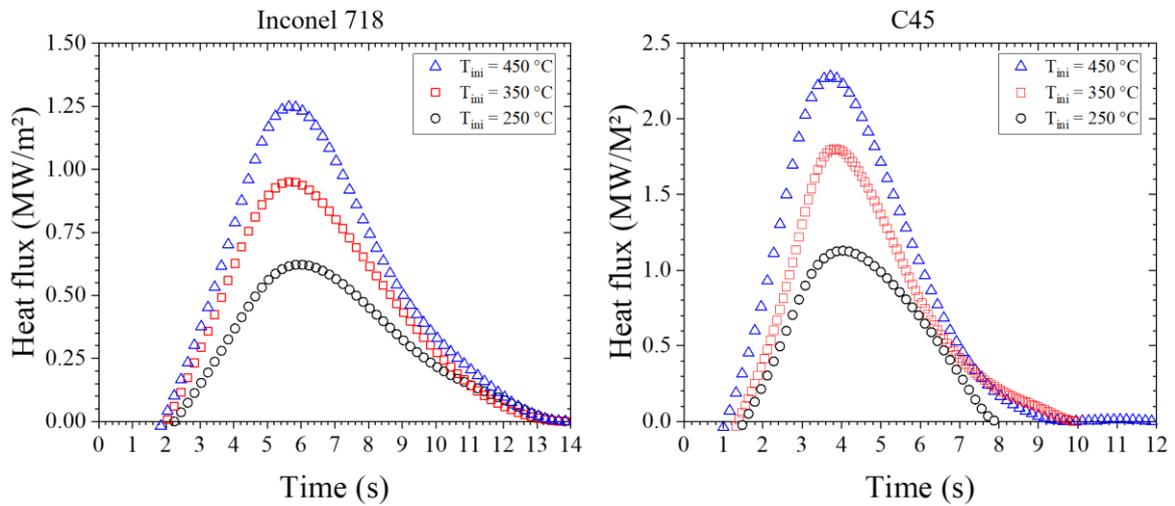

Figure 10: Influence of initial temperature on the heat flux for Inconel 718 (left) and C45 (right).

*5.3. Influence of nozzle-to-plate distance*

The influence of nozzle to plate distance is still subject of discordances in the literature [20]. In the present work, no difference in the heat flux was observed for the three different values of $H$ analysed, as can be seen in Figure 11. The same behaviour was observed by Renon et al. [21] when analysing ratios $5 < H/d_{nozzle} < 20$. A possible explanation for this result is that $H$ was not high enough to generate perturbations on the jet stream causing it to break up.

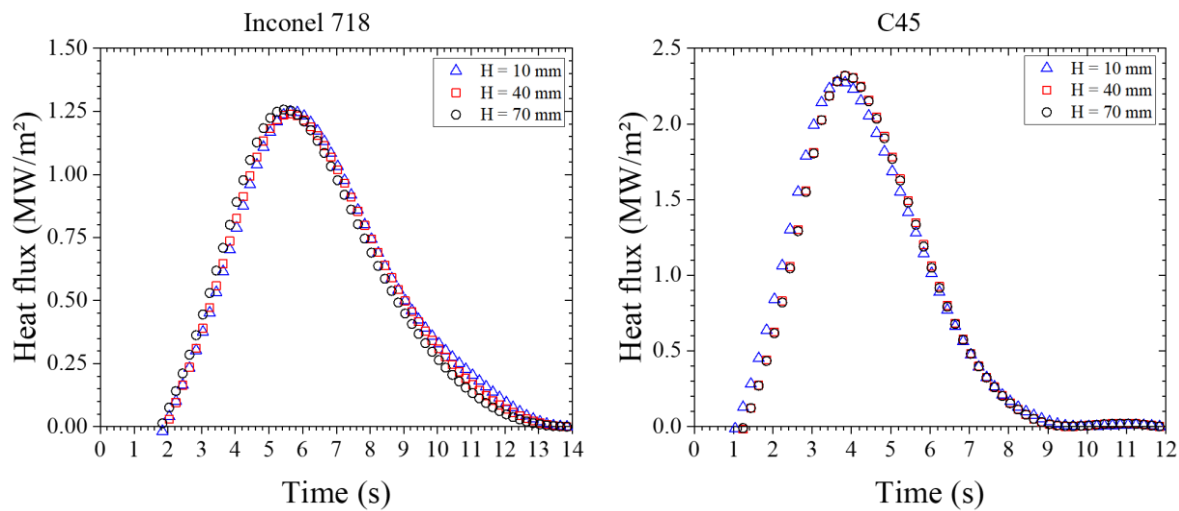

Figure 11: Influence of nozzle-to-plate distance on the heat flux for Inconel 718 (left) and C45 (right).

Besides that, for the jets analysed here, the distance $H$ is not big enough to cause any significant change in the velocity. This affirmation would not be true in the case of gas jets, in which the distance $H$ greatly affects the cooling capacity of the jet as shown by *Attalla* [22].

### 5.4. Influence of nozzle diameter and jet velocity

The influence of $d_{nozzle}$ and $V_{jet}$ is here discussed together because it is not possible to change one without changing the other when using the same flow rate. For a flow rate $Q = 4.7$ l/min, three nozzles with diameters $d_{nozzle} = 3$ mm, 5 mm, and 7 mm were analysed, resulting in velocities $V_{jet} = 2$ m/s, 4 m/s and 11 m/s, respectively. The two bigger nozzles presented approximated values for the heat flux, as shown in Figure 12. The smallest nozzle on the other hand showed enhanced heat flux for both materials, being the effect for C45 more pronounced, with an increase of approximately 30% when compared to the biggest nozzle. This result highlights the importance of high velocities in penetrating the vapour layer that may be formed in the impinging zone and also pushing away the vapour in the rewetting zone.

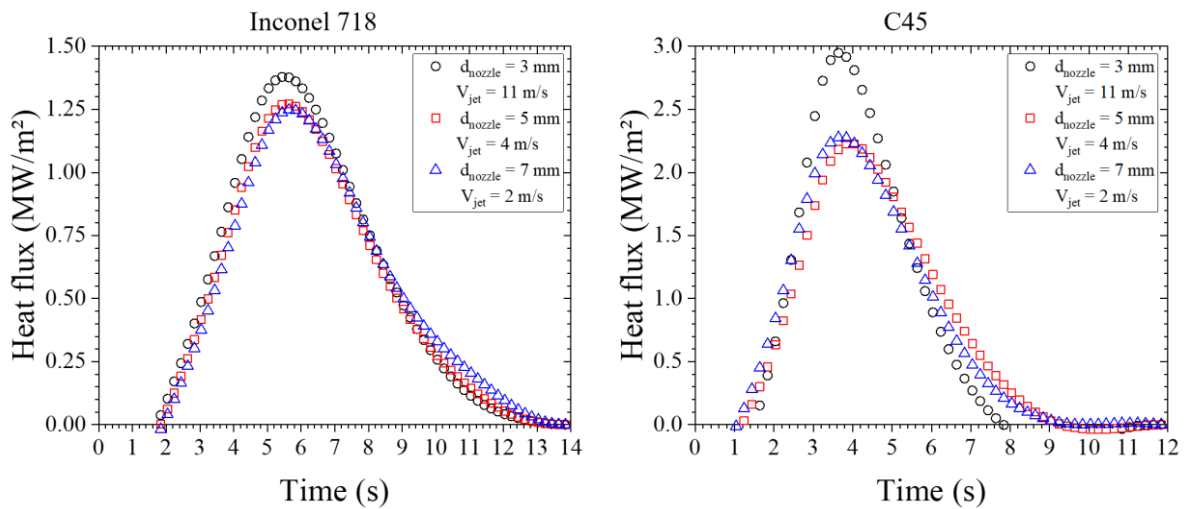

Figure 12: Influence of nozzle diameter on the heat flux for Inconel 718 (left) and C45 (right) for the same flow rate.

The influence of $d_{nozzle}$ when maintaining a constant $V_{jet}$ was not analysed here, but the consequence would be higher CHF for bigger nozzles, as showed by *Devahdhanush and Mudawar* [23]. This outcome is logically expected, because by increasing $d_{nozzle}$, the flow rate is also increased for a constant $V_{jet}$. For the same reason, the same holds true when analysing the influence of $V_{jet}$ when keeping a constant $d_{nozzle}$, as shown in Figure 13.

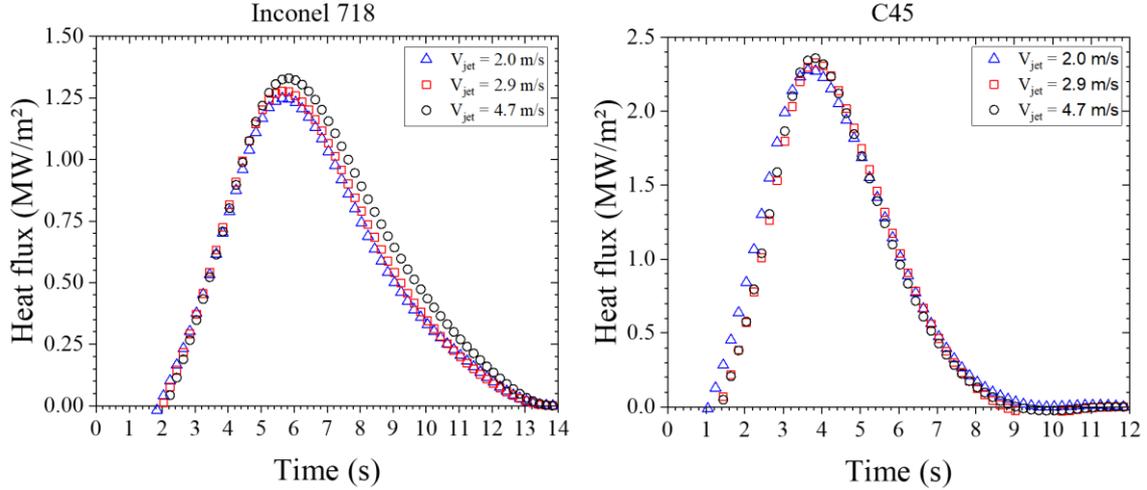

Figure 13: Influence of jet velocity on the heat flux for Inconel 718 (left) and C45 (right) for the same nozzle diameter.

The values of the velocities in Figure 13 were chosen in order to match the Re of the cases in Figure 12, as shown in Table 2. For the lowest Re (blue triangles), there is no need for comparison since the parameters $d_{nozzle}$ and $V_{jet}$ are the same. For the intermediary Re (red squares), no big differences in the heat fluxes were observed. However, for the highest Re (black circles), the higher jet velocity ($V_{jet}$ = 11 m/s) associated to the smaller diameter ($d_{nozzle}$ = 3 mm) has greater influence on the heat flux when compared to the highest jet velocity ($V_{jet}$ = 4.7 m/s) and bigger diameter ($d_{nozzle}$ = 7 mm). This effect is even greater when comparing the results for C45. If this trend continues for higher Re, these results highlight that Re, although useful, may not be the ideal parameter to be associated alone to the cooling capacity of liquid jets.

*5.5. Influence of impinging angle*

The influence of impinging angle is analysed here for three different angles $\theta_{jet}$ = 30°, 60° and 90° as shown in Figure 14. In both material plates, the highest values of heat flux were obtained for the orthogonal jet and decreased with decreasing angle. However, the difference between $\theta_{jet}$ = 90° and $\theta_{jet}$ = 60° was not so pronounced as it was for $\theta_{jet}$ = 30°. The explanation for that is in the growth and spreading of the jet over the heated surface. While the orthogonal jet impinges onto the plate and its wetting front grow radially and almost symmetrically until it covers the whole plate, as shown in Figure 15a, the same doesn't happen to the other jets. With $\theta_{jet}$ = 60°, the wetting front grows slower in the direction opposite to the jet stream, as shown in Figure 15b.

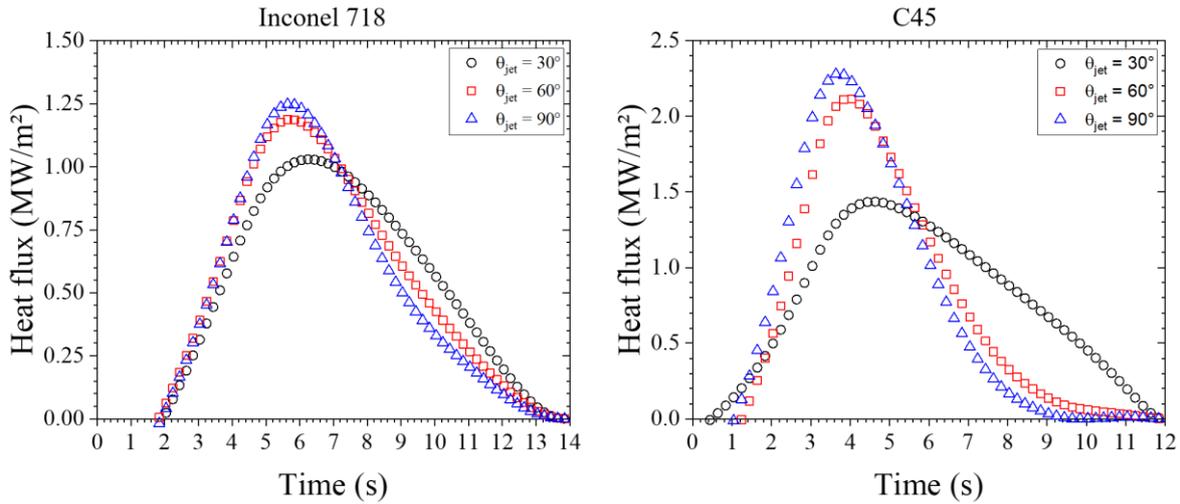

Figure 14: Influence of impinging angle on the heat flux for Inconel 718 (left) and C45 (right).

Despite the slower growth, the jet with this inclination is still able to eventually cover the whole top surface with liquid. On the other hand, the jet with $\theta_{jet} = 30°$ does not provide enough momentum to the wetting front to grow in the direction opposite to the jet stream, as shown in Figure 15c. This means that the liquid jet will not be able to cool down the plate completely with boiling mechanisms as in the other cases. This results of course in a high temperature field in the part of the plate not covered by liquid, which will be cooled eventually by combined effects of diffusion and natural convection.

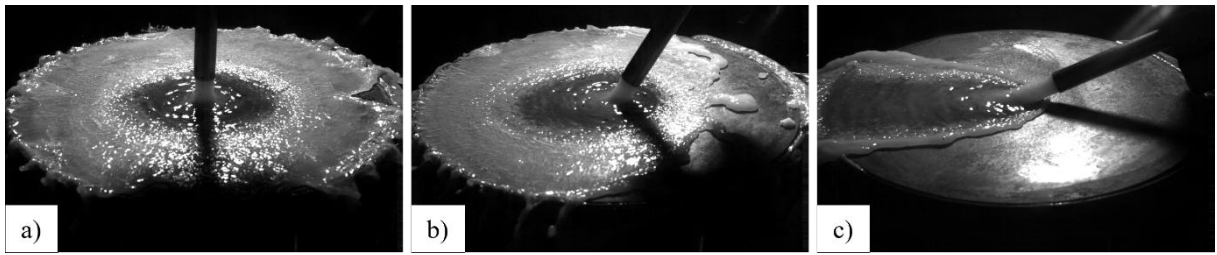

Figure 15: Influence of impingement angle on the wetting region for a) $\theta_{jet} = 90°$, b) $\theta_{jet} = 60°$, and c) $\theta_{jet} = 30°$. All images are for steel C45, emulsion with concentration $C = 8\%$, and 3 seconds after the jet touches the surface.

## 5.6. Growth of the wetting front

The growth of the wetting front is directly related to the cooling capacity of liquid jets. The thermodynamic behaviour of water and emulsion jets are shown in Figure 16a Figure 16b, respectively. Water jets presented a thinner rewetting region (transition boiling) than the emulsion jets. Another clear difference between the two liquids is the behaviour after the transition boiling. Water jets showed intense breakup and ejection of droplets from the plate. Emulsion jets, on the other hand, did not show any breakup, instead it formed a liquid film that flowed radially without presenting any signs of boiling, suggesting the presence of a vapour layer underneath it. This difference in behaviour can be due the difference in thermophysical properties between the liquid. Besides that, as shown by *Pati et al.* [24], the presence of oil can change the boiling point of the emulsion, which can be another parameter causing this effect. The growth of the wetting area, based on its diameter $d_{wet}$, is shown in Figure 16c for water, $C = 4\%$ and $C = 8\%$. The plate material in these results is Inconel 718. The concentration $C = 4\%$ presented the fastest growth, being in accordance with the heat flux analyses. Interestingly, water had an intermediary growth, suggesting that higher concentrations may reduce the growth speed.

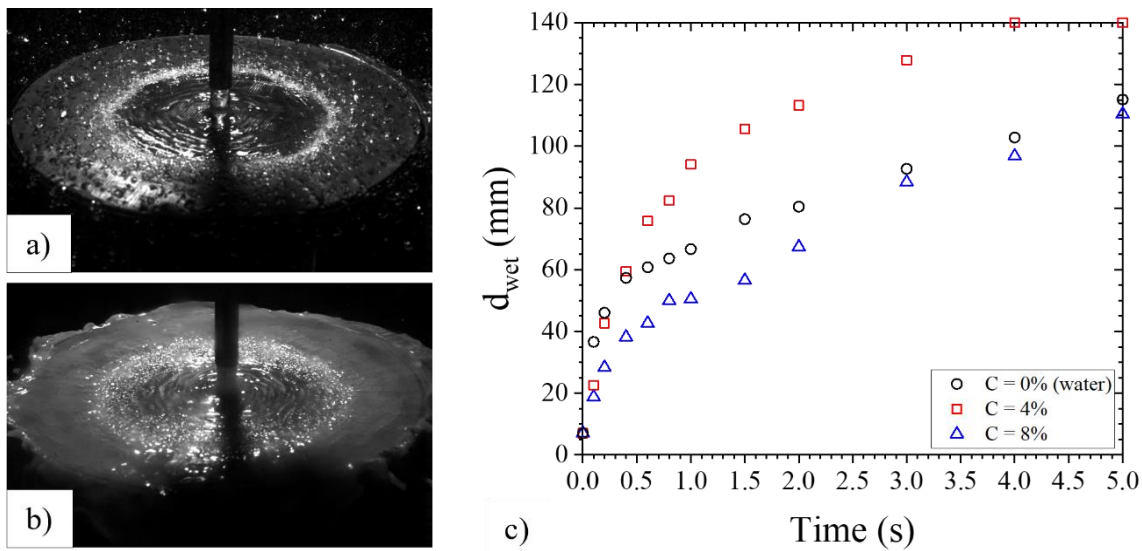

Figure 16: Difference in the thermodynamic behaviour of a) water and b) emulsion ($C = 8\%$) jets at 1.5 seconds after impinging onto an Inconel 718 plate; c) Influence of oil concentration on the growth of the wetting area (results for Inconel 718).

The influence of nozzle diameter when keeping constant flow rate is shown in Figure 17. For Inconel 718, the higher the jet velocity, the faster the growth of the wetting region in the initial stages up to t = 1.5 s. From t = 2.0 s, the jet with intermediary velocity surpasses the jet with highest velocity. For C45, the identification of the wetting region for the slower jet was not clear up to t = 1.0 s, and therefore was assumed to be equal to the nozzle diameter. After that, it grows approximately together with the jet with $V_{jet} = 4.7$ m/s. On the other hand, the highest jet velocity clearly presented the fastest growth, which agrees with the findings for the heat flux. It is also important to highlight the influence of the plate material on the growth of the wetting front. The growth was much more pronounced for the Inconel 718 plate than for the C45. To the authors' knowledge, the reason for this behaviour is not yet clear and contradicts the expected behaviour, which is C45 showing faster growth due to its higher effusivity.

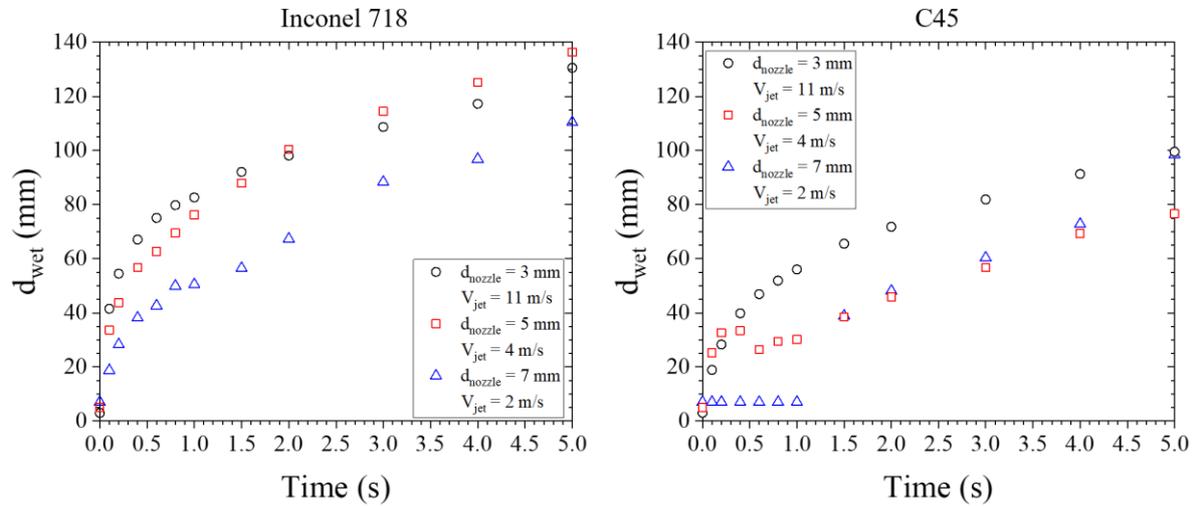

Figure 17: Influence of nozzle diameter and jet velocity on the growth of the wetting region for Inconel 718 (left) and C45 (right).

## 5.7. Dependency of CHF on location

As the plate cools down and the wetting front grows, each position of the plate passes through its respective CHF, as shown in Figure 18 for four different positions along the centreline (x = 10 mm) of the top surface and for two different liquids.

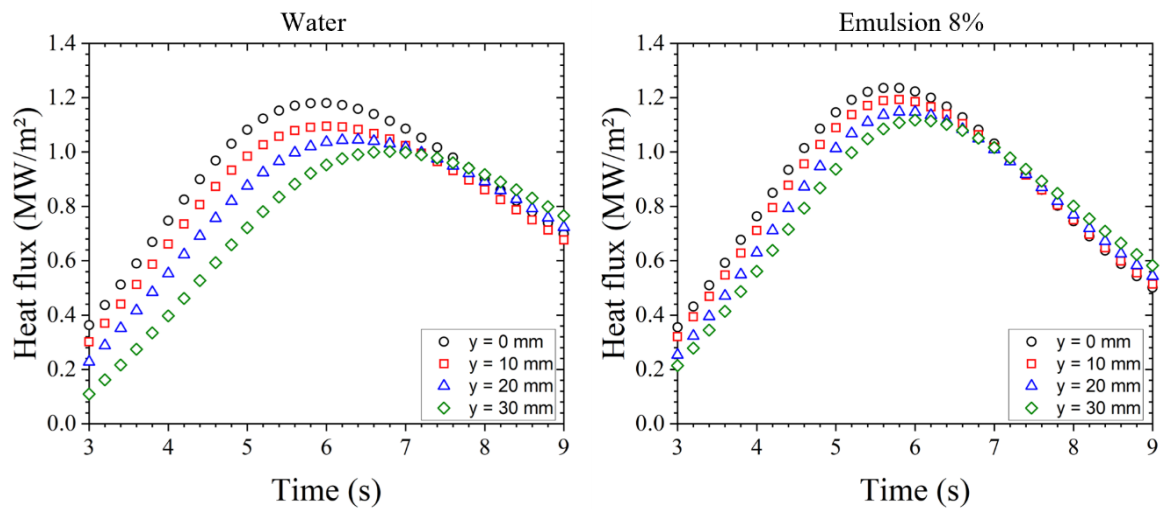

Figure 18: Heat flux for different positions along the centre axis (x = 0 mm) at the top surface and for different liquid: water (left) and emulsion with C = 8% (right). Both experiments are for Inconel 718.

For both liquids, the stagnation point presented the maximum value of CHF. The further the position is from the stagnation point, the lower is the CHF. The same behaviour was reported by *Oliveira et al.* [8]. This behaviour is influenced by many factors, for instance the lower liquid momentum and the higher liquid temperature at those locations. Another observation is that the delayed to reach the CHF is longer for more distant positions, as they need to wait for the wetting front to arrive there.

## 6. Conclusions

The cooling capacity of liquid jets and some of the parameters affecting it were investigated in this work. The cooling capacity was evaluated based on the heat flux obtained by the solution of the IHTP using the CG method implemented in the newly developed solver in OpenFoam. The experiments were done using different water and oil-in-water emulsions with different concentrations of the oil Adrana AY 401. These emulsions are commonly used in machining processes in concentrations between 4% and 10% depending on the material to be machined. Among the different concentrations analysed here, the smallest one ($C = 4\%$) presented the highest CHF, suggesting that small amounts of oil are beneficial for the cooling effect. As the concentration increases, the CHF decreases, to the point where the two higher concentrations ($C = 14\%$ and $C = 20\%$) had worse performance than pure water. Higher initial plate temperatures resulted in higher heat fluxes and showed to have great influence on the CHF. On the other hand, no significant difference was observed for the nozzle-to-plate distance ($H$) in the range analysed. The influence of nozzle diameter and jet velocity were also analysed, and the results suggest that $V_{jet}$ has a greater influence on CHF than flow rate. The impinging angle between jet and plate was also analysed. The smaller the angle, the smaller the heat flux. The reason for that was shown to be linked to the dynamic behaviour of the jet and the incapacity of very inclined jets ($\theta_{jet} = 30°$) to completely cover the surface with liquid, resulting in dry regions with high temperatures. The plate material was also shown to have significant impact on the CHF. It was observed that the ratio between the CHF for Inconel 718 and steel C45 was approximately the same as the ratio for their thermal emissivity. This was only an observation and further investigation on the subject is needed. The growth of the wetting front was shown to be faster at the early stages of impingement for jets with smaller $d_{nozzle}$ and higher $V_{jet}$ than for the opposite when using the same flow rate. An unexpected behaviour was observed when comparing the growth of the wetting region for both materials, with Inconel 718 showing faster growth. Finally, it was shown that different positions on the plate have different CHF. Regions further away from the stagnation point presented lower CHF than those closer to it.

**Acknowledgement**

The authors would like to thank the German Research Foundation (DFG) for financial support of this research. This work is related to the project "Multiphasen-Modellierungen von Kühlschmierstoff und dessen Aerosolen in der Zerspanungssimulation mit der Finite-Pointset-Methode zur Untersuchung der Wirkungsmechanismen" (439626733).

**References**

[1] Eggers, J. and Villermaux, E., 2008. Physics of liquid jets. Reports on progress in physics, 71(3), p.036601.

[2] Nukiyama, S., 1966. The maximum and minimum values of the heat Q transmitted from metal to boiling water under atmospheric pressure. International Journal of Heat and Mass Transfer, 9(12), pp.1419-1433.

[3] Kops, L. and Arenson, M., 1999. Determination of convective cooling conditions in turning. CIRP Annals, 48(1), pp.47-52.

[4] Sales, W.F., Guimaraes, G., Machado, A.R. and Ezugwu, E.O., 2002. Cooling ability of cutting fluids and measurement of the chip-tool interface temperatures. Industrial Lubrication and Tribology, 54(2), pp.57-68.

[5] Li, D., Wells, M.A., Cockcroft, S.L. and Caron, E., 2007. Effect of sample start temperature during transient boiling water heat transfer. Metallurgical and Materials Transactions B, 38, pp.901-910.

[6] Ciofalo, M., Caronia, A., Di Liberto, M. and Puleo, S., 2007. The Nukiyama curve in water spray cooling: Its derivation from temperature–time histories and its dependence on the quantities that characterize drop impact. International Journal of Heat and Mass Transfer, 50(25-26), pp.4948-4966.

[7] Gradeck, M., Ouattara, A., Maillet, D., Gardin, P. and Lebouché, M., 2011. Heat transfer associated to a hot surface quenched by a jet of oil-in-water emulsion. Experimental thermal and fluid science, 35(5), pp.841-847.

[8] Oliveira, A.V.S., Maréchal, D., Borean, J.L., Schick, V., Teixeira, J., Denis, S. and Gradeck, M., 2022. Experimental study of the heat transfer of single-jet impingement cooling onto a large heated plate near industrial conditions. International Journal of Heat and Mass Transfer, 184, p.121998..

[9] Mehdi, B., Ryll, S. and Specht, E., 2023. Analysis of the local heat transfer of quenching of moving metal sheets made of different materials using flat spray nozzles. Heat and Mass Transfer, 59(9), pp.1767-1779.

[10] Beck, J.V. and Woodbury, K.A., 2016. Inverse heat conduction problem: Sensitivity coefficient insights, filter coefficients, and intrinsic verification. International Journal of Heat and Mass Transfer, 97, pp.578-588.

[11] Duda, P., 2017. Solution of inverse heat conduction problem using the Tikhonov regularization method. Journal of Thermal Science, 26, pp.60-65.

[12] PREVODNOSTI, P.I.T., 2016. Estimation of the number of forward time steps for the sequential Beck approach used for solving inverse heat-conduction problems. Materiali in tehnologije, 50(2), pp.207-210.

[13] Xiong, P., Deng, J., Lu, T., Lu, Q., Liu, Y. and Zhang, Y., 2020. A sequential conjugate gradient method to estimate heat flux for nonlinear inverse heat conduction problem. Annals of Nuclear Energy, 149, p.107798.

[14] Hao, D.N. and Reinhardt, H.J., 1998. Gradient methods for inverse heat conduction problems. Inverse Problems in Engineering, 6(3), pp.177-211.

[15] Orlande, Helcio R.B.. Inverse Heat Transfer: Fundamentals and Applications. United States: CRC Press, 2021.

[16] Huang, C.H. and Wang, S.P., 1999. A three-dimensional inverse heat conduction problem in estimating surface heat flux by conjugate gradient method. International Journal of Heat and Mass Transfer, 42(18), pp.3387-3403.

[17] Colaco, M.J. and Orlande, H.R., 1999. Comparison of different versions of the conjugate gradient method of function estimation. Numerical Heat Transfer: Part A: Applications, 36(2), pp.229-249.

[18] Morozov, V.A., 2012. Methods for solving incorrectly posed problems. Springer Science & Business Media.

[19] Fletcher, R. and Reeves, C.M., 1964. Function minimization by conjugate gradients. The computer journal, 7(2), pp.149-154.

[20] Devahdhanush, V.S. and Mudawar, I., 2021. Review of critical heat flux (CHF) in jet impingement boiling. International Journal of Heat and Mass Transfer, 169, p.120893.

[21] Renon, C., Fénot, M., Girault, M., Guilain, S. and Assaad, B., 2021. An experimental study of local heat transfer using high Prandtl number liquid jets. International Journal of Heat and Mass Transfer, 180, p.121727.

[22] Attalla, M. and Salem, M., 2015. Experimental investigation of heat transfer for a jet impinging obliquely on a flat surface. Experimental Heat Transfer, 28(4), pp.378-391.

[23] Devahdhanush, V.S. and Mudawar, I., 2021. Critical heat flux of confined round single jet and jet array impingement boiling. International Journal of Heat and Mass Transfer, 169, p.120857.

[24] Pati, A.R., Mandal, S., Dash, A., Barik, K., Munshi, B. and Mohapatra, S.S., 2018. Oil-in-water emulsion spray: A novel methodology for the enhancement of heat transfer rate in film boiling regime. International Communications in Heat and Mass Transfer, 98, pp.96-105.